\documentclass[twocolumn,aps,secnumarabic%
,showkeys,amsmath,amssymb,superscriptaddress]{revtex4}

\usepackage{graphicx}
\usepackage{dcolumn}
\usepackage{bm}

\begin{document}


\title{First principles study of density, viscosity, and diffusion
coefficients of liquid MgSiO$_3$ at conditions of the Earth's deep
mantle}

\author{Jones T. K. Wan}
\affiliation{Department of Physics, Hong Kong
University of Science and Technology, Clear Water Bay, Kowloon,
Hong Kong SAR}
\affiliation{Department of Chemistry and Princeton Materials
Institute, Princeton University, Princeton, New Jersey 08544, USA}

\author{Thomas S. Duffy}
\affiliation{Department of Geoscience, Princeton University,
Princeton, New Jersey 08544, USA}

\author{Sandro Scandolo}
 \affiliation{The Adbus Salam International Centre for Theoretical
Physics and INFM/Democritos National Simulation Center, \\
Strada Costiera 11, 34100 Trieste, Italy}

\author{Roberto Car}
\affiliation{Department of Chemistry and Princeton Materials
Institute, Princeton University, Princeton, New Jersey 08544, USA}

\date{\today}

\begin{abstract}
Constant-pressure constant-temperature {\it ab initio} molecular
dynamics simulations at high temperatures have been used to study
MgSiO$_3$ liquid, the major constituent of the Earth's lower mantle
to conditions of the Earth's core-mantle boundary (CMB).
We have performed variable-cell
{\it ab initio} molecular dynamic simulations at relevant
thermodynamic conditions across one of the measured melting
curves. The calculated equilibrium volumes and densities are
compared with the simulations using an orthorhombic perovskite
configuration under the same conditions. For molten MgSiO$_3$, we
have determined the diffusion coefficients and shear viscosities
at different thermodynamic conditions. Our results provide new
constraints on the properties of molten MgSiO$_3$ at conditions
near the core-mantle boundary.  The volume change on fusion is
positive throughout the pressure-temperature conditions examined
and ranges from 5\% at 88 GPa and 3500 K to 2.9\% at 120 GPa and
5000 K. Nevertheless, neutral or negatively buoyant melts from
(Mg,Fe)SiO$_3$ perovskite compositions at deep lower mantle conditions are
consistent with existing experimental constraints on solid-liquid
partition coefficients for Fe. Our simulations indicate that
MgSiO$_3$ is liquid at 120 GPa and 4500 K, consistent with the
lower  range of experimental melting curves for this material.
Linear extrapolation of our results indicates that the densities
of liquid and solid perovskite MgSiO$_3$ will become equal near 180 GPa.
\end{abstract}

\maketitle

\section{Introduction}
Melting is a ubiquitous process in planetary interiors and one of
the dominant mechanisms for thermal transport and chemical
differentiation in planets.  The properties of
silicate liquids are thus essential for understanding a wide range of
geophysical phenomena related to the deep Earth and its origin and
evolution.  (Mg, Fe)SiO$_3$ perovskite is the most abundant
mineral in the Earth's deep mantle.   Here we report the first
{\it ab initio} simulation of the properties of liquid MgSiO$_3$
at conditions corresponding to the deep interior of the Earth.

There are a number of lines of evidence that strongly indicate the Earth (and
other terrestrial planets)  were partially or wholly molten at least at
certain intervals during the accretion process
~\citep{stevenson_1989,solomatov_2000}.  The subsequent cooling and
crystallization of a magma ocean may lead to chemical
differentiation of the mantle~\citep{ohtani_1985}.  Thus,
understanding the dynamics of a terrestrial magma ocean is
essential to understanding the initial conditions for the thermal
and chemical evolution of the Earth. The liquid viscosity and its depth
dependence  are among  the important parameters that characterize an early
(liquid-rich) magma ocean~\citep{solomatov_2000} and melt/crystal density
inversions could greatly modify the structure and cooling history of such an
ocean~\citep{solomatov_2000}. At present, only low-pressure
experimental data or calculations based on empirical
potentials~\citep{wasserman_1993} have been used to constrain these
models.

The existence of small degrees ($<1\%$)  of
partial melt in the present lower mantle has been suggested as an explanation
for the strong relative variations of shear velocity relative to compressional
($\partial\ln V_S/\partial\ln V_P$) velocity in the deep
mantle~\citep{duffy_1992}. In the D" region at the base of the mantle, a
seismic ultra-low velocity zone (ULVZ) of thickness 5-40 km has been detected
locally on top of the core-mantle
boundary~\citep{garnero_1993,revenaugh_1997,helmberger_2000}. These regions are characterized
by seismic compressional and shear velocity reductions of $\sim$ 10\% and
$\sim$ 30\% respectively.  The presence of relatively large degrees of partial
melt ($\sim$ 5-30\%) has been proposed as the most plausible explanation for
these features~\citep{williams_1996}. There is also evidence locally for
smaller shear velocity reductions in D" plausibly consistent with lesser
amounts of partial melt at depths as great as 300 km above the core mantle
boundary~\citep{wen_2001}.

The presence of melt in the deep mantle can greatly affect a
number of physical properties of the region.  The viscosity of the
melt may strongly modify heat transport and convective circulation
within the boundary layer~\citep{williams_1996}.  The presence of a
partially molten layer suggests the  density contrast between
solid and melt should be small. The buoyancy of the melt will be
controlled by the intrinsic density difference, as well as
compositional differences  (i.e., Fe enrichment) between the melt
and solid.  Density inversions between silicate melts and
equilibrium liquidus crystals have been extensively studied at
upper mantle conditions~\citep{agee_1998,rigden_1984}. However,
there are only limited constraints on possible density inversions
under conditions of the Earth's lower mantle. Recently, it has 
been proposed based on laboratory measurements to 15 GPa that
basaltic melts may become denser than mantle peridotite at
conditions near the base of the mantle~\citep{ohtani_2001}.
However,  these results are subject to considerable uncertainty
due to the long extrapolations involved. More direct determination
of physical properties of the components of deep mantle melts are
needed.

Recent developments in the atomistic simulation of solids and liquids
based on the full solution of the quantum mechanical equations for
the electrons allow the theoretical study from
first principles, i.e.  without empirical or adjustable parameters,
of the structural, thermal, and elastic properties of minerals
at arbitrary conditions of pressure and temperature.
Recent works have addressed successfully the thermoelastic properties
of solid MgSiO$_3$ perovskite at Earth's mantle conditions
~\citep{karki_am_1997,karki_prb_2000,oganov_epsl_2001,oganov_nature_2001%
,wentzcovitch_prl_2004}.
Here we focus on the properties of liquid MgSiO$_3$ at high pressures,
which we study using first-principles molecular dynamics at constant
pressure. We determine liquid densities and compare them with solid
perovskite densities at similar conditions to extract melting properties. We
also calculate or estimate dynamical properties of liquid MgSiO$_3$ such
as diffusion and viscosity.

\section{Technical details}
{\it Ab initio} molecular dynamic simulations~\citep{car_1985} are
performed using density functional
theory~\citep{hohenberg_1964,kohn_1965} within the generalized
gradient approximation~\citep{perdew_1996} (GGA). We use ultrasoft
pseudopotential~\citep{vanderbilt_1990} for O and norm-conserving
pseudopotentials~\citep{troullier_1991} for Mg and Si. Nonlinear
core corrections~\citep{louie_1982} were used for Mg. Kohn-Sham
orbitals were expanded in plane waves with a kinetic energy cut
off of 30 Ry. The simulation cell contained 80 atoms (16 MgSiO$_3$
units) and a time step of 13 a.t.u. was used.
Variable-cell dynamics were used to impose the required
value of the pressure~\citep{parrinello_1980}. Temperature was controlled by
a Nos\'{e} thermostat~\citep{nose_1984,hoover_1985}.
The liquid configuration was obtained by heating the system to 10000 K and
then decreasing the temperature to 5000 K in a total time of 1 ps.

The diffusivity, or diffusion coefficient $(D)$ is estimated by
calculating the mean square displacement of the atoms as a
function of time. The diffusion coefficient is given by the
Einstein relation~\citep{allen_1987}:
\begin{equation}
D\approx\lim_{t\rightarrow\infty}{1\over 6t}\langle\vert{\bf
r}(t_0+t)-{\bf r}(t_0)\vert^2\rangle
=\lim_{t\rightarrow\infty}{1\over 6t}\langle\delta R^2(t)\rangle.
\end{equation}
The calculated mean square displacement were fitted by a
straight line: $\langle\delta R^2(t)\rangle\approx6Dt+b$.
The viscosity $(\eta)$ is estimated using the generalized
Debye-Stokes-Einstein formula derived by
Zwanzig~\citep{zwanzig_jcp_1983}:
\begin{equation}
{D\eta\over k_B T n^{1/3}}=0.0658(2+\eta/\eta_l)=C_l.
\end{equation}
Here $n$ is the number density of the atoms, $n=N/V$. The constant
$C_l$ depends on the ratio of the shear viscosity $(\eta)$ to the
longitudinal viscosity $(\eta_l)$. Although these numbers are not
actually available, $C_l$ has bounds that can vary between 0.132
and 0.181. In our work, we adopt a typical value of $C_l=0.171$.
Wasserman {\it et al}~\citep{wasserman_1993}
studied the transport properties of
perovskite melts by molecular dynamics under temperatures
(3500-6000 K) and pressures (5.1-78 GPa) and their results agreed
reasonably well with Zwanzig's formula. The general validity of
Zwanzig's formula has been studied in detail in other
systems~\citep{march_1999,bagchi_2001,gezelter_1999}.

\section{Simulation results}
\subsection{Materials properties at high pressures and temperatures}
\begin{figure}
\includegraphics[width=0.8\columnwidth]{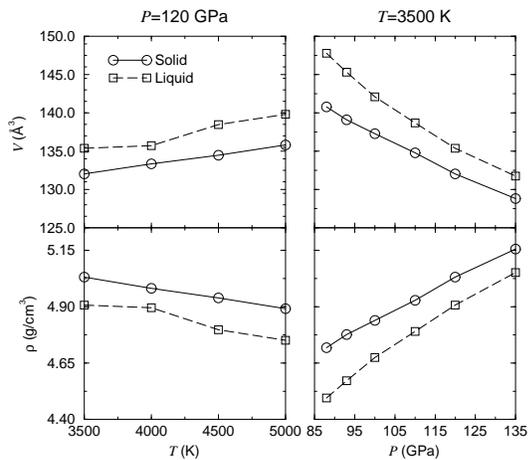}
\caption{\label{vol_3500_120G}(Left panels) Volumes and densities
of solid and molten MgSiO$_3$ at different temperatures, pressure
is fixed at 120 GPa. Note the difference between $T>$ 4500 K and
$T<$ 4000 K. (Right panels) The same quantities but with
temperature fixed at 3500 K and pressure changes. There is a
tendency of density inversion between melting solid and liquid at
higher pressures.}
\end{figure}

The calculated equilibrium volumes and densities of four MgSiO$_3$ units at
3500 K as a function of pressure are given in Table~\ref{prop_3500K}. The
results are shown in the right panels of Fig.~\ref{vol_3500_120G}.  Our
results at 88 GPa with the liquid configuration showed that atoms are
diffusive with a diffusion coefficient of $1.14\times 10^{-5}$ cm$^2$/s. Even
though this thermodynamic point (3500 K and 88 GPa) is close to the melting
line obtained by the Berkeley group~\citep{jeanloz_1996,knittle_1989,heinz_1987}, the
stability of the liquid in our simulations does not allow us to conclude that
3500 K is above the melting line of the theoretical model employed. Liquid
(meta)stability in the few picoseconds of a first-principles simulation could
be due to super-cooling effects~\citep{allen_1987}.  We did not estimate the
diffusion coefficients at higher pressures because all the available melting
curves suggest that MgSiO$_3$ can only be solid at these thermodynamic
conditions.
\begin{table}
\caption{\label{prop_3500K} Calculated  properties of
MgSiO$_3$ at 3500 K  as a function of pressure using solid perovskite and
liquid configurations. For liquid configuration at 88 GPa, atoms
are diffusive and the estimated diffusion coefficient is $1.14 \times 10^{-5}$cm$^2$/s.}
\begin{tabular}{ccccccccc}
  \hline
  &  &\multicolumn{2}{c}{Solid perovskite}&\multicolumn{3}{c}{Liquid} \\
   $T$ & $P$  & $V_S$ & $\rho_S$  & $V_L$ & $\rho_L$ & $\Delta V$\\
   (K) & (GPa) & ($\AA^3$) & (g/cm$^3$) & ($\AA^3$) & (g/cm$^3$) & (\%)\\ \hline
  3500 &  88 & 140.787 & 4.718 & 147.764 & 4.495 & 4.96\\
  3500 &  93 & 139.076 & 4.776 & 145.296 & 4.571 & 4.47\\
  3500 & 100 & 137.263 & 4.839 & 142.098 & 4.674 & 3.52\\
  3500 & 110 & 134.786 & 4.928 & 138.681 & 4.790 & 2.89\\
  3500 & 120 & 132.013 & 5.031 & 135.360 & 4.907 & 2.54\\
  3500 & 135 & 128.815 & 5.156 & 131.762 & 5.041 & 2.29\\ \hline
\end{tabular}
\end{table}

Regarding the simulations with liquid configuration, when pressure increases,
the atoms approach each other and become no longer diffusive. This can be
observed by our estimated diffusion coefficients at 88 GPa ($1.14\times
10^{-5}$ cm$^2$/s) and 120 GPa ($0.04\times 10^{-5}$ cm$^2$/s).  In other
words, the under-cooled liquid is glassified. This agrees with the melting
curve obtained from 
Knittle and Jeanloz~\citep{knittle_1989}, 
and Heniz and Jeanloz~\citep{heinz_1987}.
In
addition, as pressure increases, the percentage difference of volumes between
solid and liquid MgSiO$_3$ becomes smaller, which indicate a tendency of
density inversion at higher pressures.  In the left panels of
Fig.~\ref{vol_3500_120G}, the volumes and densities are shown as a function of
temperature up to 5000 K.  Pressure is fixed at 120 GPa. This is close to that
at the top of the D" region in the mantle.  This pressure is also close to
where MgSiO$_3$ transforms to a post-perovskite phase
~\citep{murakami_science_2004} In this study, our solid state calculations are
restricted to the perovskite crystal structure. The volume of the
post-perovskite (CaIrO$_3$-type) phase is $\sim$1-1.5\% less than perovskite
~\citep{oganov_nature_2004,tsuchiya_epsl_2004,shieh_agu_2004} at deep lower
mantle pressures, so this transformation will increase the density contrast
between solid and liquid.  Because of uncertainties associated with the
Clapeyron slope of the transition and the deep mantle geotherm, it is possible
that at the base of the mantle the geotherm will cross back into the
perovskite stability field ~\citep{hernlund_agu_2004}.

Although melting has not been observed during the simulations, our results
provide an upper bound of the solid volume. The numerical results are given in
Table~\ref{prop_120G}. For liquid simulations, the changes of volume and
density can be approximately divided into two regions, say, $T>$ 4500 K and
$T<$ 4000 K. It may be due to the possible glassification of liquid below 4000
K. Similar observation of glassification can be found if one looks at the mean
square displacements of the liquid at different temperatures.

\subsection{Diffusion near the core-mantle boundary}\label{diff}
\begin{figure}
\includegraphics[width=0.8\columnwidth]{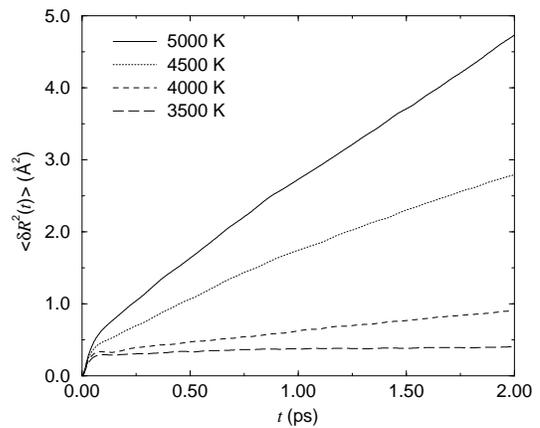}
\caption{\label{msd_liq}Mean square displacements of molten
MgSiO$_3$ at 120 GPa. The liquid starts to glassify at temperature
around 4000 K.}
\end{figure}

In Table~\ref{prop_120G}, we tabulate the equilibrium volumes and
densities of solid and molten MgSiO$_3$, together with the
estimated diffusion coefficients and viscosities of the molten
MgSiO$_3$ as a function of temperature. Pressure is fixed at 120
GPa. The volumes and densities are shown in the left panels of
Fig.~\ref{vol_3500_120G}.
The estimated diffusion coefficient drops drastically as the
temperature drops from 4500 K to 4000 K. This is a strong
indication of glassification of molten MgSiO$_3$. The mean square
displacements of molten MgSiO$_3$ at different temperatures are
shown in Fig.~\ref{msd_liq}.
At higher temperatures
(4500 K $\sim$ 5000 K), The atoms are clearly diffusive. This
provides strong evidence that molten MgSiO$_3$ exists at 4500 K
and 120 GPa.
\begin{table*}
\caption{\label{prop_120G} Calculated materials properties of
MgSiO$_3$ at 120 GPa as a function of temperature using solid perovskite and
liquid configurations.}
\begin{tabular}{cccccccc}
  \hline
  &  &\multicolumn{2}{c}{Solid}&\multicolumn{4}{c}{Liquid} \\
   $T$ & $P$  & $V_S$ & $\rho_S$  & $V_L$ & $\rho_L$ & $D$ & $\eta$ \\
   (K) & (GPa) & ($\AA^3$) & (g/cm$^3$) & ($\AA^3$) & (g/cm$^3$) & (10$^{-5}$ cm$^2$/s) & (cp)\\ \hline
  3500 & 120 & 132.013 & 5.031 & 135.360 & 4.907 & 0.04 & - \\
  4000 & 120 & 133.340 & 4.981 & 135.712 & 4.895 & 0.46 &107.74\\
  4500 & 120 & 134.476 & 4.939 & 138.448 & 4.798 & 1.77 & 31.46\\
  5000 & 120 & 135.792 & 4.891 & 139.795 & 4.751 & 3.33 & 18.51\\ \hline
\end{tabular}
\end{table*}
\begin{figure}
\includegraphics[width=0.8\columnwidth]{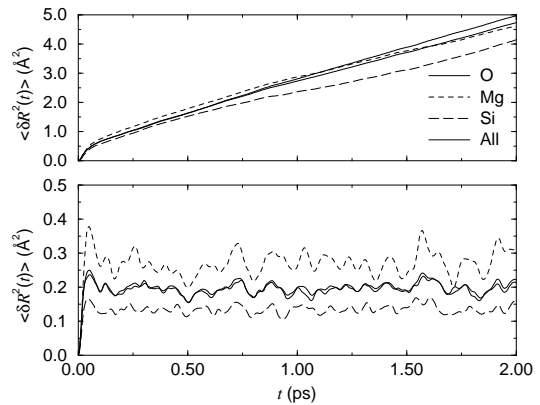}
\caption{\label{msd_5000_120G}Mean square displacements of molten
(upper panel) and solid (lower panel) MgSiO$_3$ at 5000 K and 120
GPa. Solid lines: O atoms, dashed lines: Mg atoms, long dashed
lines: Si atoms, solid bold lines: all atoms.}
\end{figure}

The mean square displacements of different species of atoms are shown in
Fig.~\ref{msd_5000_120G}. Both solid and liquid results are shown. In the
solid, Si atoms are bounded by the octahedra formed by the O atoms, and each
octahedron is then surrounded by 8 Mg atoms. Hence Mg atoms are expected to
have a larger vibrational amplitude then that of the other species, and Si
atoms are expected to have the smallest vibrational amplitude
(Fig.~\ref{msd_5000_120G}).  In molten MgSiO$_3$, the slopes of the mean
square displacements of each species is close to that of each other.  This
suggests the self-diffusion rate of each species is almost the same.  The
simplest explanation is that the three types of atoms diffuse like a single
molecule, which has to be examined carefully.  However, as the diffusion
mechanism is beyond the scope of this work, it will be the subject of further
studies.

\section{Discussion and conclusion}
At 3500 K, the density of crystalline MgSiO$_3$ in the perovskite structure is
5\% greater than the liquid at 88 GPa, and this density excess is 2.9\% at 120
GPa and 5000 K.  The transformation to a post-perovskite phase would further
enhance this density contrast.  Thus, we find that liquid MgSiO$_3$ is always
less dense than the corresponding solid under conditions encountered in Earth'
s mantle. This implies that stability of melts in the deep lower mantle
requires in addition  compositional differences between melt and solid.  It is
well known on the basis of mineral-melt partitioning data for perovskite that
Fe is preferentially partitioned into the melt. Hence, in addition to the
volume of fusion, Fe enrichment in the melt must be considered when evaluating
the stability of melts in an idealized (Mg,Fe)SiO$_3$ lower
mantle~\citep{knittle_1998}.

The effect of Fe on densities in crystalline and liquid
(Mg,Fe)SiO$_3$ perovskites can be calculated using known
thermoelastic parameters~\citep{jackson_1998,deschamps_2003} and
assuming the effect of Fe on density in the liquid is similar to
its effect on the solid. For solid (Mg,Fe)SiO$_3$ compositions
with Mg/(Mg+Fe) ratios of 0.90 to 0.95, the solid-liquid Fe
partition coefficients required to be in equilibrium with
neutrally buoyant melt at 120 GPa and 5000 K range from  0.31 to
0.48. These are within the wide range of reported experimental
partition coefficients for perovskite compositions at lower mantle
pressures (0.12-0.57)~\citep{ito_1987,mcfarlane_1994,knittle_1998}.
Thus, the combination of the small, positive volume change of
fusion determined here with Fe partitioning constraints from
experiment suggest that negative or neutrally buoyant melts are
plausible in the D"  and ULVZ layers at least with respect to
perovskite structure solids.
Partition coefficients
within the experimentally measured range (0.2-0.34) also produce
neutrally buoyant melts at 88 GPa and 3500 K.    Thus, these
results suggest that small amounts of neutrally buoyant partial
melt in the deep lower mantle could also contribute to the
anomalous $\partial\ln V_S/\partial\ln V_P$ observed in this
region~\citep{duffy_1992}. Negative buoyancy under these conditions
could instead lead to melt pooling in D" and the
ULVZ~\citep{knittle_1998}. Better constraints on Fe partition
coefficients under deep mantle conditions are needed to
distinguish between these possibilities.

Extrapolation of the trend in Fig.~\ref{vol_3500_120G} indicates
that MgSiO$_3$ liquid and crystal densities will become equal near
180 GPa. There are few experimental constraints on densities in
MgSiO$_3$ liquids under these extreme conditions.  The density
change along the Hugoniot in Mg$_2$SiO$_4$ upon shock melting at
150 GPa is small and there is a suggestion of enhanced
compressibility in the melt~\citep{brown_1987}.
More recent shock experiments on MgSiO$_3$ compositions have suggested that
the density of the melt becomes comparable to that of the solid near 120 GPa,
and at 170 GPa the melt density exceeds the solid density
~\citep{akins_grl_2004}.  However, the shock data near 120 GPa (which are
directly comparable to our simulations) have uncertainties that allow melt
densities to be as much as  several percent less than the solid, and thus they
are not necessarily inconsistent with the present results.

The viscosities of silicate liquids have been investigated
experimentally using a wide variety of techniques at ambient and
low pressures~\citep{doremus_2002}.  However, there are few
constraints on the behavior of silica poor and relatively
depolymerized melts that are relevant to melting in the deep
Earth.  The viscosity of a diopside (CaMgSi$_2$O$_6$) liquid has
recently been determined to high pressures and temperatures using
the falling sphere method~\citep{reid_2003}. At 8-13 GPa and
2200-2470 K,  the reported viscosities range from 28-510
centipoise which are generally larger than our higher pressure and
temperature values (Table~\ref{prop_120G}). The temperature
dependence of viscosity for MgSiO$_3$ liquid remains strong even
at 120 GPa (Table~\ref{prop_120G}).

In this work, we preformed {\it ab initio} molecular dynamic
simulations to study MgSiO$_3$, the main mineral composition in the Earth's
lower mantle. By applying variable-cell dynamics, we studied both
solid perovskite and molten MgSiO$_3$ in constant $NPT$ ensemble without the
need for optimizing the cell parameters. At thermodynamic
conditions close to that of the core-mantle boundary, we have
constrained the viscosity, diffusion coefficient, and  density of
liquid MgSiOi$_3$. Our results also support the validity of the
relatively low melting curve for MgSiO$_3$
perovskite~\citep{jeanloz_1996,knittle_1989,heinz_1987} as compared
with that reported by
Zerr and Boehler~\citep{zerr_1993}.

The next step is to study the diffusion mechanism in molten
MgSiO$_3$ as well as the chemical
heterogeneity on the diffusion and melting behavior. With the
increasing computation power, it is anticipated that the
applications of {\it ab initio} molecular dynamics can contribute
to detailed understanding of mineral physics at the core-mantle
boundary.

\begin{acknowledgments}
J.T.K. Wan acknowledges the financial support from The Croucher
Foundation. This work was supported by the NSF. We thank A. Kubo
for valuable discussion.
\end{acknowledgments}



\end{document}